\documentstyle[12pt,epsf]{article}

\catcode`\@=11
\long\def\@makefntext#1{
\protect\noindent \hbox to 3.2pt {\hskip-.9pt
$^{{\ninerm\@thefnmark}}$\hfil}#1\hfill}

\def\thefootnote{\fnsymbol{footnote}}
\def\@makefnmark{\hbox to 0pt{$^{\@thefnmark}$\hss}}

\def\ps@myheadings{\let\@mkboth\@gobbletwo
\def\@oddhead{\hbox{}
\rightmark\hfil\ninerm\thepage}
\def\@oddfoot{}\def\@evenhead{\ninerm\thepage\hfil
\leftmark\hbox{}}\def\@evenfoot{}
\def\sectionmark##1{}\def\subsectionmark##1{}}

\renewcommand{\thefootnote}{\fnsymbol{footnote}}

\newcounter{sectionc}\newcounter{subsectionc}\newcounter{subsubsectionc}
\renewcommand{\section}[1] {\vspace*{0.6cm}\addtocounter{sectionc}{1}
\setcounter{subsectionc}{0}\setcounter{subsubsectionc}{0}\noindent
	{\bf\thesectionc. #1}\par\vspace*{0.4cm}}
\renewcommand{\subsection}[1] {\vspace*{0.6cm}\addtocounter{subsectionc}{1}
	\setcounter{subsubsectionc}{0}\noindent
	{\it\thesectionc.\thesubsectionc. #1}\par\vspace*{0.4cm}}
\renewcommand{\subsubsection}[1]{\vspace*{0.6cm}\addtocounter{subsubsectionc}{1}
	\noindent {\rm\thesectionc.\thesubsectionc.\thesubsubsectionc.
	#1}\par\vspace*{0.4cm}}

\newcounter{appendixc}
\newcounter{subappendixc}[appendixc]
\newcounter{subsubappendixc}[subappendixc]

\renewcommand{\appendix}[1] {\vspace*{0.6cm}
        \refstepcounter{appendixc}
        \setcounter{figure}{0}
        \setcounter{table}{0}
        \setcounter{equation}{0}
        \renewcommand{\thefigure}{\Alph{appendixc}.\arabic{figure}}
        \renewcommand{\thetable}{\Alph{appendixc}.\arabic{table}}
        \renewcommand{\theappendixc}{\Alph{appendixc}}
        \renewcommand{\theequation}{\Alph{appendixc}.\arabic{equation}}
        \noindent{\bf Appendix \theappendixc #1}\par\vspace*{0.4cm}}

\def\abstracts#1{{

\centering{\begin{minipage}{12.2truecm}\footnotesize\baselineskip=12pt\noindent
	\centerline{\footnotesize ABSTRACT}\vspace*{0.3cm}
	\parindent=0pt #1
	\end{minipage}}\par}}

\renewenvironment{thebibliography}[1]
	{\begin{list}{\arabic{enumi}.}
	{\usecounter{enumi}\setlength{\parsep}{0pt}
\setlength{\leftmargin 1.25cm}{\rightmargin 0pt}
	 \setlength{\itemsep}{0pt} \settowidth
	{\labelwidth}{#1.}\sloppy}}{\end{list}}

\newcounter{itemlistc}
\newcounter{romanlistc}
\newcounter{alphlistc}
\newcounter{arabiclistc}

\newcommand{\fcaption}[1]{
        \refstepcounter{figure}
        \setbox\@tempboxa = \hbox{\footnotesize Fig.~\thefigure. #1}
        \ifdim \wd\@tempboxa > 6in
           {\begin{center}
        \parbox{6in}{\footnotesize\baselineskip=12pt Fig.~\thefigure. #1}
            \end{center}}
        \else
             {\begin{center}
             {\footnotesize Fig.~\thefigure. #1}
              \end{center}}
        \fi}

\newcommand{\tcaption}[1]{
        \refstepcounter{table}
        \setbox\@tempboxa = \hbox{\footnotesize Table~\thetable. #1}
        \ifdim \wd\@tempboxa > 6in
           {\begin{center}
        \parbox{6in}{\footnotesize\baselineskip=12pt Table~\thetable. #1}
            \end{center}}
        \else
             {\begin{center}
             {\footnotesize Table~\thetable. #1}
              \end{center}}
        \fi}

\def\@citex[#1]#2{\if@filesw\immediate\write\@auxout
	{\string\citation{#2}}\fi
\def\@citea{}\@cite{\@for\@citeb:=#2\do
	{\@citea\def\@citea{,}\@ifundefined
	{b@\@citeb}{{\bf ?}\@warning
	{Citation `\@citeb' on page \thepage \space undefined}}
	{\csname b@\@citeb\endcsname}}}{#1}}

\newif\if@cghi
\def\cite{\@cghitrue\@ifnextchar [{\@tempswatrue
	\@citex}{\@tempswafalse\@citex[]}}
\def\citelow{\@cghifalse\@ifnextchar [{\@tempswatrue
	\@citex}{\@tempswafalse\@citex[]}}
\def\@cite#1#2{{$\null^{#1}$\if@tempswa\typeout
	{IJCGA warning: optional citation argument
	ignored: `#2'} \fi}}

\font\ninerm=cmr9

\newcommand{\beq}{\begin{equation}}
\newcommand{\eeq}{\end{equation}}
\newcommand{\beqa}{\begin{eqnarray}}
\newcommand{\eeqa}{\end{eqnarray}}

\newcommand{\NPB}[1]{{\it Nucl. Phys.}\ {\bf B{#1}}}
\newcommand{\PLB}[1]{{\it Phys. Lett.}\ {\bf B{#1}}}

\newcommand{\lae}{\begin{array}{c}\,\sim\vspace{-21pt}\\< \end{array}}
\newcommand{\gae}{\begin{array}{c}\,\sim\vspace{-21pt}\\> \end{array}}

\begin{document}

\setcounter{footnote}{0}
\renewcommand{\thefootnote}{\fnsymbol{footnote}}

\centerline{\normalsize\bf ISOSPIN BREAKING AND THE TOP-QUARK MASS IN}
\baselineskip=22pt
\centerline{\normalsize\bf MODELS OF DYNAMICAL ELECTROWEAK SYMMETRY
BREAKING\footnote{Talk presented by R.~S.~Chivukula at the
{\it Workshop on Top Quark Physics}, Iowa State University,
Ames, IA, May 25-26, 1995  and the {\it Yukawa International Seminar
`95}, Yukawa Institute, Kyoto, Aug. 21-25, 1995}}
\baselineskip=16pt

\centerline{\footnotesize R.S. Chivukula, B.A. Dobrescu, and J. Terning}
\baselineskip=13pt
\centerline{\footnotesize\it Physics Department, Boston University}
\baselineskip=12pt
\centerline{\footnotesize\it 590 Commonwealth Ave.}
\centerline{\footnotesize\it Boston, MA 02215 USA}
\centerline{\footnotesize E-mail: sekhar@bu.edu,
dobrescu@budoe.bu.edu, terning@calvin.bu.edu}
\baselineskip=16pt
\centerline{\footnotesize\sf BUHEP-95-22 \& hep-ph/9506450}

\vspace*{0.9cm}
 \abstracts{ In this talk we review the physics of top-quark mass
generation in models of dynamical electroweak symmetry breaking and
the constraints on this physics arising from limits on the deviation
of the weak interaction $\rho$-parameter from one. We then discuss
top-color assisted technicolor in this context.  }

\normalsize\baselineskip=15pt
\setcounter{footnote}{0}
\renewcommand{\thefootnote}{\alph{footnote}}

\section{$m_t$ in Models of Dynamical Electroweak Symmetry Breaking}
\label{sec:mt}

In technicolor models, the masses of the ordinary fermions are due to
their coupling, through additional, broken, extended-technicolor (ETC)
gauge-interactions\cite{ETC}, to the technifermions:

\beq
\epsfxsize 4cm \centerline{\epsffile{tETC.eps}}
\eeq

\noindent
which leads to a mass for the top-quark

\beq
m_t \approx {g^2 \over M^2_{ETC}} \langle\bar{U} U \rangle_{M_{ETC}}\ ,
\eeq

\noindent
where we have been careful to note that it is the value of the
technifermion condensate renormalized at the scale $M_{ETC}$ which is
relevant.

For a QCD-like technicolor, where there is no substantial
difference between $\langle\bar{U} U \rangle_{M_{ETC}}$ and
$\langle\bar{U} U \rangle_{\Lambda_{TC}}$, we can use naive
dimensional analysis\cite{dimanal} to estimate the technifermion
condensate, arriving at a top-quark mass

\beq
m_t \approx {g^2 \over M^2_{ETC}} 4 \pi F^3\ ,
\eeq

\noindent
where $F$ is the analog of the pion decay-constant (93 MeV, in my
normalization) in QCD. We can invert this relation to find the
characteristic mass-scale of top-quark mass-generation

\beq
{M_{ETC} \over g} \approx 1\ {\rm TeV} \left({F  \over 246\ {\rm
GeV}}\right)^{3\over 2} \left({175\ {\rm GeV} \over m_t}\right)^{1\over
2}.
\label{blitz}
\eeq

We immediately see that the scale of top-quark mass generation is
likely to be {\it quite} low, unless the value of the technifermion
condensate ($\langle\bar{U} U \rangle_{M_{ETC}}$) can be raised
significantly above the value predicted by naive dimensional
analysis. The prospect of such a low ETC-scale is both tantalizing and
problematic. As we will see in the next section, constraints from the
deviation of the weak interaction $\rho$ parameter from one suggest
that the scale may have to be larger than one TeV.

\begin{figure}[htb]
\vspace{0.25cm}
\epsfxsize 8cm \centerline{\epsffile{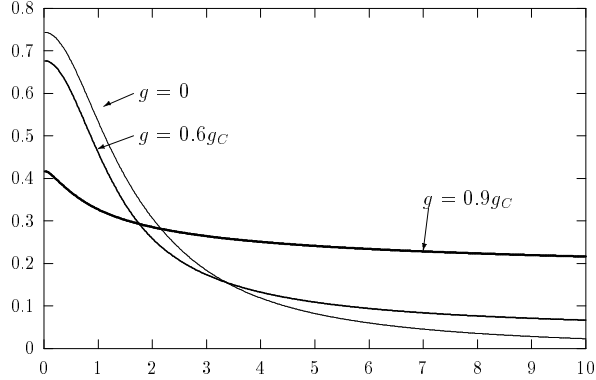}}
\caption{Plot\protect\cite{evans} of technifermion self energy $\Sigma(k)$ vs.
momentum
(both in TeV), as predicted by the gap-equation in the rainbow
approximation, for various strengths of the ETC coupling relative to
their critical value $g_C$.}
\label{figone}
\vspace{0.25cm}
\end{figure}

The most promising approach\cite{strongETC} to enhance the
technifermion condensate and accommodate the top-quark mass is
``strong-ETC''. As the size of the ETC-coupling
approaches the critical value for chiral symmetry breaking, it is
possible to enhance the running technifermion self-energy $\Sigma(k)$
at large momenta (see Figure \ref{figone}).

\noindent
Since the technifermion condensate is related to the trace of the
fermion propagator.

\beq
\langle\bar{U} U \rangle_{M_{ETC}} \propto \int^{M^2_{ETC}}
dk^2 \Sigma(k)\ ,
\eeq

\noindent
a slowly-falling running-mass translates to an enhanced
condensate\footnote{More physically, in terms of the relevant
low-energy theory, it can be shown that the enhancement of the
top-quark mass is due to the dynamical generation of a light scalar
state\cite{Comp}}.

\begin{figure}[htb]
\vspace{0.25cm}
\epsfxsize 8cm \centerline{\epsffile{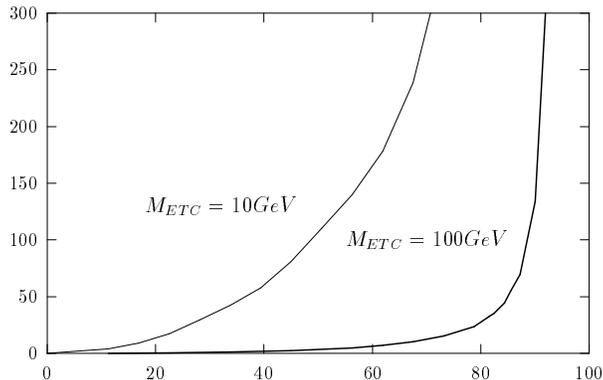}}
\caption{Plot\protect\cite{evans} of top mass (in GeV) vs. ETC
coupling ($g/g_c$ in \%), as predicted by gap-equation in the rainbow
approximation, for ETC scales of 10 and 100 TeV.}
\label{figtwo}
\vspace{0.25cm}
\end{figure}

Unfortunately, there is no such thing as a free lunch.  As we see from
Fig. \ref{figtwo}, the enhancement of the technifermion self-energy in
strong-ETC theories comes at the cost of a ``fine-tuning" of the
strength of the ETC coupling relative to the critical value where the
ETC interactions would, in and of themselves, generate chiral symmetry
breaking. In the context of the NJL approximation\cite{NJL}, we find that
enhancement of the top quark mass is directly related to the severity
of this adjustment,

\beq
{\langle\bar{U} U \rangle_{\Lambda_{TC}}\over \langle\bar{U} U
\rangle_{M_{ETC}}} \approx {\Delta g^2 \over g^2_c}~,
\label{price}
\eeq

\noindent
where $\Delta g^2 \equiv g^2 - g^2_C$.

\section{$\Delta\rho_*$}
\label{sec:deltarho}

The physics which is responsible for top-quark mass generation must
violate custodial $SU(2)$ since, after all, this physics must give rise
to the disparate top- and bottom-quark masses.

\subsection{Direct Contributions}

ETC operators which violate custodial isospin by two units ($\Delta I
= 2$) are particularly dangerous\cite{isospin2}. Denoting the
right-handed technifermion doublet by $\psi_R$, consider the operator

\beq
{g^2 \over M^2} \left(\bar{\Psi}_R \gamma_\mu \sigma_3 \Psi_R\right)^2~~,
\label{cohler}
\eeq

\noindent
which can be interpreted as the (mass-)mixing of the $Z$ with an isosinglet ETC
gauge-boson

\beq
\epsfxsize 5cm \centerline{\epsffile{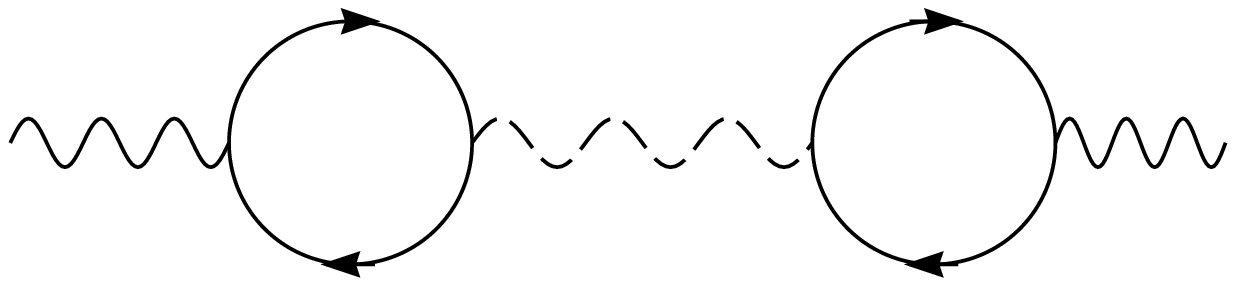}}
\eeq

\noindent
and give rise to a contribution to $\Delta\rho_*$\footnote{
Contributions of this sort occur naturally in ETC-models which give
rise to the top-quark mass\protect\cite{wu}.}.

If there are $N_D$ doublets of the technifermions and they
give rise to a contribution to $M^2_W$ proportional to $N_D F^2$, the
contribution of the operator in eqn. (\ref{cohler}) to the $\rho$
parameter can be calculated to be

\beqa
\Delta\rho_* & \approx &  {2 g^2 \over M^2} {N^2_D F^4 \over v^2} \\
& \approx &  12\%\  g^2 \left({N_D F^2 \over (246 \ {\rm
GeV})^2}\right)^2 \left({1 \ {\rm TeV} \over M}\right)^2 ~.
\label{important}
\eeqa

\noindent
Current limits\cite{tc2isospin} on the parameter $T$ ($\Delta
\rho_* = \alpha T$) imply that $\Delta\rho_* \lae 0.4\% $.

There are two ways in which one may try to satisfy this constraint.
The equation above implies

\beq
{M\over g} \gae  5.5\ {\rm TeV} \left( {N_D F^2 \over (246\ {\rm
GeV})^2}\right)\ .
\eeq

\noindent
If $N_D F^2 \approx (246\ {\rm GeV})^2$, that is if the sector giving
rise to the top-quark mass is responsible for the bulk of EWSB, then
the scale $M$ must be much larger than the naive 1 TeV expectation in
QCD-like technicolor.  Comparing this with eqns. (\ref{blitz}) and
(\ref{price}) above, we see that the enhancement of the condensate
needed requires a fine-tuning of order 3\%
($\approx (1/5.5)^2$) in
order to produce a top-quark mass of order 175 GeV.
Alternatively, we may re-write the bound as

\beq
F \lae {105\ {\rm GeV} \over \sqrt{N_D}} \left( {M/g \over 1\ {\rm
TeV}}\right)^{1\over 2}
\eeq

\noindent
If $M/g$ is of order 1 TeV, it is necessary that the sector
responsible for top quark mass generation {\em not} give rise to the
bulk of EWSB.  This is essentially what happens in multiscale models
\cite{multi,chiraltc} and in top-color assisted technicolor
\cite{top-color}.

\subsection{Indirect Contributions}

A second class of potentially dangerous contributions come from
isospin violation in the technifermion mass spectrum.  In a manner
analogous to the contribution\cite{Drho} of the $t-b$ mass splitting to $\Delta
\rho$, any difference in the dynamical masses of two technifermions in
the {\em same} doublet will give rise to deviations in the $\rho$
parameter from one. The size of this effect can be estimated \`a la
Pagels-Stokar\cite{pagels},

\beq
\Delta \rho_* \propto {N_D\, d \over 16 \pi^2}
\left({\Sigma_U(0) - \Sigma_D(0)}\over v\right)^2~~,
\eeq

\noindent
where $N_D$ and $d$ are the the number of doublets and dimension of
the technicolor representation respectively.  Since we require $\Delta
\rho_* \lae 0.4$\%, the equation above implies

\beq
N_D\, d \left({\Delta\Sigma(0)\over m_t}\right)^2 \lae 1.3 \, \, .
\label{ilimit}
\eeq

\noindent
{}From this we see that, $\Delta \Sigma(0)$ must be less than of order
$m_t$ \footnote{Perhaps, given the crude approximations involved, one may be
able to live with $d=2$ in the fundamental of and $SU(2)$ technicolor
group with one doublet.}.

However, if the $t$ and $b$ get their mass from the same technidoublet, then at
the ETC-scale we expect that there is no difference between the $t$,
$b$ and the corresponding technifermions\cite{strongETC}

\beqa
\Delta\Sigma(M_{ETC})& \equiv & \Sigma_U(M_{ETC}) - \Sigma_D(M_{ETC})
\approx  \nonumber \\
\Delta m(M_{ETC}) & \equiv & m_t(M_{ETC}) - m_b(M_{ETC})~.
\label{problem}
\eeqa

\noindent
Furthermore, if QCD is the only interaction which contributes to the
scaling of the $t$ and $b$ masses, we expect $\Delta m(M_{ETC}) \approx
m^{pole}_t$, and from scaling properties of the technifermion
self-energies, we expect $\Delta\Sigma(0) \gae \Delta
\Sigma(M_{ETC})$.

There are two ways to avoid these constraints.  One is that perhaps
there are {\it additional} interactions which contribute to the
scaling of the top- and bottom-masses below the ETC scale, and hence
that $\Delta m(M_{ETC}) \ll m_t^{pole}$. This would be the case if the
$t$ and/or $b$ get only a {\em portion} of their mass from the
technicolor interactions, and would imply that the third generation
must have (strong) interactions different from the technifermions (and
possibly from the first and second generations).  Another possibility
is that the $t$ and $b$ get mass from {\em different} technidoublets,
each of which have isospin-symmetric masses. The first alternative is
the solution chosen in top-color assisted technicolor models (see
below), while the latter has only recently begun to be
explored\cite{isosymmetric}.

\section{Case Study: Top-Color Assisted Technicolor}
\label{sec:tcii}

Recently\cite{top-color}, Hill has introduced a model in which a
top-condensate is driven by the combination of a strong, but
spontaneously broken and non-confining, isospin-symmetric top-color
interaction and an additional (either weak or strong) isospin-breaking
$U(1)$ interaction which couple only to the third generation
quarks.
At low-energies, the top-color and hypercharge interactions of the
third generation quarks may be approximated by four-fermion operators

\beq
{\cal L}_{4f} = -{{4\pi
\kappa_{tc}}\over{M^2}}\left[\overline{\psi}\gamma_\mu {{\lambda^a}\over{2}}
\psi \right]^2
-{{4\pi \kappa_1}\over{M^2}}\left[{1\over3}\overline{\psi_L}\gamma_\mu  \psi_L
+{4\over3}\overline{t_R}\gamma_\mu  t_R
-{2\over3}\overline{b_R}\gamma_\mu  b_R
\right]^2~,
\label{L4t}
\eeq

\noindent
where $\psi$ represents the top-bottom doublet, $\kappa_{tc}$ and
$\kappa_1$ are related respectively to the top-color and $U(1)$
gauge-couplings squared, and where (for convenience) we have assumed
that the top-color and $U(1)$ gauge-boson masses are comparable and of
order $M$. In order to produce a large top quark mass without giving
rise to a correspondingly large bottom quark mass, the combination of
the top-color and extra hypercharge interactions are assumed to be
critical in the case of the top quark but not the bottom quark:

\beq
\kappa^t_{eff} = \kappa_{tc} +{1\over3}\kappa_1 >
\kappa_c = {{3\pi}\over{8}} >
\kappa^b_{eff}=\kappa_{tc} -{1\over 6}\kappa_1~.
\label{kc}
\eeq

The contribution of the top-color sector to electroweak symmetry
breaking can be quantified by the $F$-constant of this sector.  In the
NJL approximation\cite{NJL}, for $M$ of order 1 TeV, and $m_t
\approx 175$ GeV, we find
\beq
f_t^2 \equiv    {{N_c }\over{8\pi^2}}\, m_t^2
\log\left({{M^2}\over{m_t^2}}\right) \approx (64\ {\rm GeV})^2~.
\label{ft}
\eeq

\subsection{Direct Isospin Violation}

As $f_t$ is small compared to 246 GeV,
technicolor is necessary to
produce the bulk of EWSB and to give mass to the light
fermions. However, the heavy and light fermions must mix --- hence, we
would naturally expect that at least some of the {\it technifermions}
carry the extra $U(1)$ interaction.  If the additional $U(1)$
interactions violate custodial symmetry\footnote{It has been
noted\cite{isosymmetric} that if the top- and bottom-quarks receive
their masses from {\em different} technidoublets, it is possible to
assign the extra $U(1)$ quantum numbers in a custodially invariant
fashion.}, the $U(1)$ coupling will have to be quite small to keep
this contribution to $\Delta \rho_*$ small\cite{tc2isospin}.  We will
illustrate this in the one-family technicolor\cite{onefam} model,
assuming that techniquarks and technileptons carry $U(1)$-charges
proportional to the hypercharge of the corresponding ordinary
fermion\footnote{Note that this choice is anomaly-free.}. We can
rewrite the effective $U(1)$ interaction of the technifermions as

\beq
{\cal L}_{4T1} =-{{4\pi
\kappa_1}\over{M^2}}\left[{1\over3}\overline{\Psi}\gamma_\mu  \Psi
+\overline{\Psi}_R\gamma_\mu  \sigma^3 \Psi_R
-\overline{L} \gamma_\mu L
+ \overline{L}_R \gamma_\mu\sigma^3 L_R
\right]^2~,
\label{L4T}
\eeq

\noindent
where $\Psi$ and $L$ are the techniquark and technilepton
doublets respectively.

{}From the analysis given above (eqn. (\ref{important})), we see that
the contribution to $\Delta
\rho_*$ is\cite{tc2isospin}:
\beq
\Delta \rho_{*}^{\rm T} \approx 152\% \
\kappa_1 \left({{1\ {\rm TeV}}\over{M}}\right)^2~.
\label{rhoT}
\eeq

\noindent
Therefore if $M$ is of order 1 TeV, and the extra $U(1)$ has
isospin-violating couplings to technifermions, $\kappa_1$ must be
extremely small.

\subsection{Indirect Isospin Violation}

In principle, since the isospin-splitting of the top and bottom are
driven by the combination of top-color and the extra $U(1)$, the
technifermions can be degenerate. In this case, the only indirect
contribution to the $\rho$ parameter at one-loop is the usual
contribution coming from loops of top- and bottom-quarks\cite{Drho}.
However, since there are additional interactions felt by the
third-generation of quarks, there are ``two-loop'' contributions of
the form

\beq
\epsfxsize 5cm \centerline{\epsffile{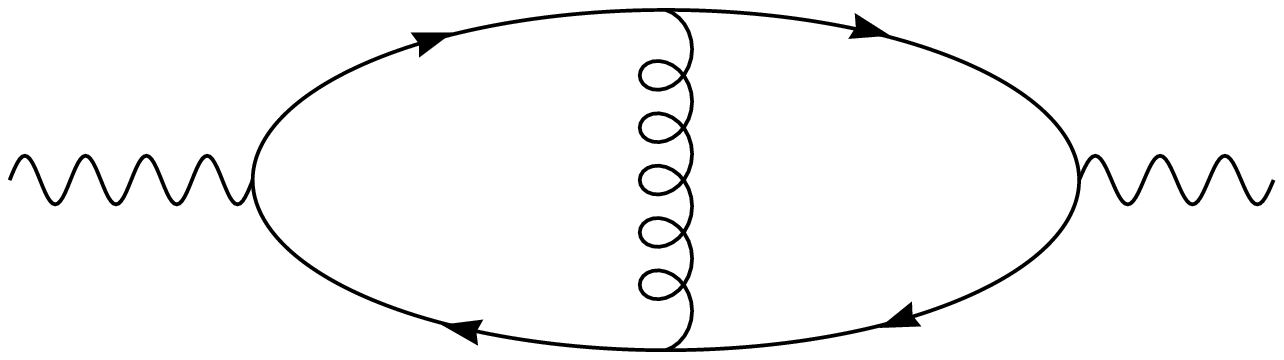}}
\label{twoloop}
\eeq
This contribution yields\cite{tc2isospin}
\beq
\Delta \rho_{*}^{\rm tc} \approx 0.53\%
\left({\kappa_{tc}\over \kappa_c}\right)
\left(1\ {\rm TeV}\over M\right)^2
\left(f_t \over 64\ {\rm GeV}\right)^4
{}~.
\label{rhot}
\eeq

\noindent
Combining this with eqn. (\ref{ft}), we find that
\beq
M \gae 1.4\ {\rm TeV}
\eeq

\noindent
for $\kappa_{tc} \approx \kappa_c$.  This immediately puts a
constraint on the mass of the top-color gluon which is comparable to
the direct limits currently obtained by CDF\cite{cdftwojet}.

\subsection{Fine-Tuning}

Finally, we must require that the sum of the effects of eqns. (\ref{rhoT})
and (\ref{rhot}) do not give rise to an experimentally disallowed
contribution to the $\rho$ parameter.  Equation (\ref{rhoT}) implies
that $\kappa_1$ must either be very small, or $M$ very large. However,
we must also simultaneously satisfy the constraint of eqn. (\ref{kc}),
which implies that

\beq
{\Delta\kappa_{tc}\over\kappa_c} = \left|{{\kappa_{tc}-\kappa_c}
\over \kappa_c}\right| \le
{1 \over 3} {\kappa_1\over \kappa_c}~,
\label{dktc}
\eeq

\begin{figure}[htb]
\vspace{0.25cm}
\epsfxsize 8cm \centerline{\epsffile{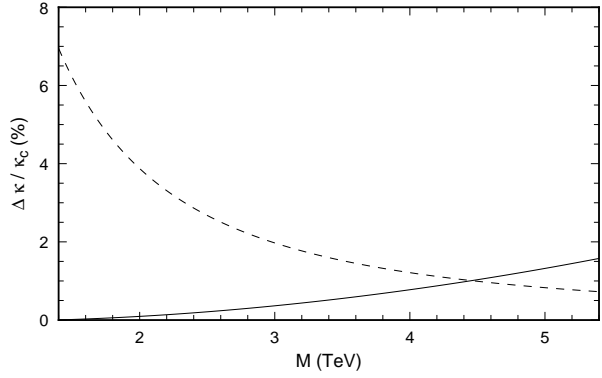}}
\caption{The amount of fine-tuning required\protect\cite{tc2isospin} in the
${\rm TC}^2$ model.
The dashed line is the amount of fine-tuning in $\Delta\kappa_{eff}$
required to keep $m_t$ much lighter than $M$, see equation
(\ref{dkefft}). The solid curve shows the amount of fine-tuning (see
equation (\ref{dktc})) in $\Delta\kappa_{tc}$ required to satisfy the
bound $\Delta \rho_* < 0.4$\%.  The region excluded by the
experimental constraint on $\Delta \rho_*$ is above the solid
curve.}
\label{figfour}
\vspace{0.25cm}
\end{figure}

\noindent
Therefore, if $M$ is low and $\kappa_1$ is small, the
top-color coupling must be tuned close to the critical value for
chiral symmetry breaking. On the other hand, if $\kappa_1$ is not
small and $M$ is relatively large the {\em total} coupling of the
top-quark must be tuned close to the critical NJL value\cite{NJL} for
chiral symmetry breaking in order to keep the top-quark mass low,
\beq
{\Delta\kappa_{eff}\over\kappa_c} =
{{\kappa^t_{eff}-\kappa_c}\over \kappa_c} =
{{{m^2_t\over M^2}\log{M^2\over m^2_t}} \over {1-{{m^2_t\over
M^2}\log{M^2\over m^2_t}}}}~.
\label{dkefft}
\eeq
These two constraints are shown in Fig. \ref{figfour}.  For $M>$ 1.4
TeV, we find that either $\Delta\kappa_{tc}/
\kappa_c$ or $\Delta\kappa_{eff}/\kappa_c$ must be tuned to
less than 1\%.  This trade-off in fine tunings is displayed in
figure 4.  For the ``best" case where both tunings are of order 1\%,
$M=4.5$ TeV.

\section{Conclusions}
\label{sec:concl}

We have seen that a large top quark mass has a number of
important implications for dynamical electroweak symmetry
breaking:

\begin{itemize}
\item[$\bullet$]
A large top-quark mass naturally implies, in models of dynamical
electroweak symmetry breaking, the possibility of a correspondingly
low scale for the scale of top flavor-physics.

\item[$\bullet$]
The physics responsible for the large isospin breaking in the $t-b$
mass splitting can lead to potentially dangerous ``direct" and ``indirect"
effects in the $W$ and $Z$ masses.

\item[$\bullet$]
The direct and indirect effects can be mitigated if the sector which
is responsible for the top- and bottom-masses does {\em not} provide the
bulk of electroweak symmetry breaking and, conversely, if the sector
responsible for the $W$ and $Z$ masses gives rise to only a {\em small
portion} of the top- and bottom-masses. This can happen only if the top
and bottom feel {\em strong} interactions which are not shared by the
technifermions and, possibly, the first two generations.

\item[$\bullet$]
In top-color assisted technicolor, the extra top-color interactions
give rise to additional indirect contributions to $\Delta \rho$, and
we must require that $M_{g} \gae 1.4\  {\rm TeV}$. Furthermore, If the extra
$U(1)$ has isospin-violating couplings to technifermions, we require
fine-tuning of order 1\%.

\end{itemize}


\centerline{\bf Acknowledgments}

We thank Tom Appelquist, Nick Evans, and Ken Lane for helpful
conversations, and Mike Dugan for help in preparing the manuscript.
R.S.C. acknowledges the support of an NSF Presidential Young
Investigator Award, and a DOE Outstanding Junior Investigator Award.
{\em This work was supported in part by the National Science
Foundation under grant PHY-9057173, and by the Department of Energy
under grant DE-FG02-91ER40676.}



\begin{thebibliography}{99}
\frenchspacing

\bibitem{ETC} S. Dimopoulos and L. Susskind, \NPB{155}
(1979) 237;
E. Eichten and K. Lane, \PLB{90} (1980) 125.

\bibitem{dimanal}A. Manohar and H. Georgi, \NPB{234} (1984) 189.

\bibitem{strongETC}
T. Appelquist, M. Einhorn, T. Takeuchi, and L.C.R. Wijewardhana,
{\em Phys. Lett.} {\bf 220B}, 223 (1989);
V.A. Miransky and K. Yamawaki, {\em Mod. Phys. Lett.} {\bf A4}, 129 (1989);
K. Matumoto {\em Prog. Theor. Phys. Lett.} {\bf 81}, 277 (1989) .

\bibitem{evans} N. Evans, \PLB{331} (1994) 378.

\bibitem{Comp}R.S.~Chivukula, K.~Lane, and A.G.~Cohen, {\em Nucl. Phys.} {\bf
B 343}, 554 (1990);
T. Appelquist, J. Terning, and L. Wijewardhana, {\em Phys. Rev.} {\bf 44},
871 (1991).

\bibitem{NJL} Y. Nambu and G. Jona-Lasinio, {\em Phys. Rev.} {\bf 122}
(1961) 345.

\bibitem{isospin2}T. Appelquist et al., {\em Phys. Rev.} {\bf D31}
(1985) 1676.

\bibitem{wu} See, for example, G.-H. Wu, these proceedings and hep-ph/9412206.

\bibitem{top-color}C.T. Hill, these proceedings and  \PLB{345} (1995) 483.

\bibitem{onefam}E. Farhi and L. Susskind, {\em Phys. Rev.} {\bf D20}
(1979) 3404.

\bibitem{tc2isospin}R.S. Chivukula,  B.A. Dobrescu, and
J. Terning, hep-ph/9503203.

\bibitem{multi}K. Lane and E. Eichten, {\em Phys. Lett.} {\bf B222}
(1989) 274.

\bibitem{chiraltc}J. Terning, hep-ph/9410233,
{\em Phys. Lett.} {\bf B344} (1995) 279.

\bibitem{Drho}M. Einhorn, D. Jones, and M. Veltman,
{\em Nucl. Phys.} {\bf B191} (1981) 146.

\bibitem{pagels} H. Pagels and S. Stokar, {\em Phys. Rev.} {\bf D20}
(1979) 2947; B. Holdom, \PLB{226} (1989) 137.

\bibitem{isosymmetric}K. Lane and E. Eichten, hep-ph/9503433.

\bibitem{cdftwojet} A. Beretvas, these proceedings, and F. Abe
{\it et. al.}, CDF collaboration, FERMILAB-PUB-94/405-E.

\end{thebibliography}
\end{document}